%% file: main.tex
\documentclass[sigconf]{acmart} 

\usepackage{etoolbox}
\usepackage{ifthen}
\usepackage{comment} 
\usepackage{float}
\usepackage{graphicx} 
\usepackage{makecell} 
\usepackage{makecell,booktabs,array}
\newcolumntype{L}{>{\raggedright\arraybackslash}X} 
\newcolumntype{Y}{>{\raggedleft\arraybackslash}X}  
\newboolean{makeXTRA}
\setboolean{makeXTRA}{false} 

\makeatletter%
\newboolean{maketitleAfterAllMeta}
\@ifclassloaded{acmart}%
{\setboolean{maketitleAfterAllMeta}{true}}%
{\setboolean{maketitleAfterAllMeta}{false}}%
\makeatother%

\usepackage[utf8]{inputenc}
\usepackage[T1]{fontenc}

\usepackage{listings}
\usepackage{xcolor}

\newcommand{\xspeed}[1]{\ensuremath{#1\times}}  
\usepackage{microtype}

\usepackage{anyfontsize}
\usepackage[10pt]{moresize}

\usepackage{amsmath}
\usepackage{amsfonts}
\usepackage{mathtools}

\makeatletter%
\@ifclassloaded{acmart}{}{\usepackage{amssymb}}
\makeatother%

\usepackage{siunitx}
\usepackage{xfrac}
\usepackage{tabu}
\usepackage{xspace}

\makeatletter%
\@ifclassloaded{acmart}{}%
{\usepackage[usenames,dvipsnames,table]{xcolor}}
\makeatother%

\makeatletter%
\@ifclassloaded{acmart}{}
{\@ifclassloaded{elsarticle}{}
{\usepackage{cite}}}
\makeatother%

\usepackage{setspace}

\usepackage{verbatim}

\usepackage{booktabs}
\usepackage{makecell}
\usepackage{multicol}
\usepackage{multirow}
\usepackage{tabularx}
\usepackage{array}

\usepackage{stfloats}

\makeatletter
\@ifclassloaded{IEEEtran}{%
  \let\MYcaption\@makecaption%
  \usepackage[font=footnotesize]{subcaption}%
  \let\@makecaption\MYcaption%
}{%
  \usepackage{subcaption}%
}%
\makeatother

\usepackage[textsize=tiny]{todonotes}

\makeatletter%
\@ifclassloaded{IEEEtran}{}{}%
\usepackage{enumitem}
\makeatother%

\usepackage{quoting}

\usepackage{listings}

\makeatletter%
\@ifclassloaded{acmart}%
{}{}%
\@ifclassloaded{elsarticle}%
{}{}
\@ifclassloaded{llncs}%
{}{}
\@ifclassloaded{jpconf}%
{%
  \usepackage{doi}%
}{}
\@ifclassloaded{IEEEtran}%
{%
  
  \newcommand{\ifIEEEtran}[1]{#1}
}%
{%
  \newcommand{\ifIEEEtran}[1]{}
  \usepackage{flushend}
}%
\makeatother%

\makeatletter%
\@ifpackageloaded{hyperref}%
{}%
{%
  \usepackage[hyphens]{url}
  \usepackage[breaklinks=true,bookmarks=false]{hyperref}
  \hypersetup{
    colorlinks=false,%
    pdfborder = 0 0 0
  }
}%
\makeatother%

\usepackage[capitalize,noabbrev,capitalize]{cleveref} 

\usepackage{stackengine}
\usepackage{graphicx}
\usepackage{tikz}
\usepackage{pgfplots}
\usepackage{pgfplotstable}

\ifthenelse{\boolean{makeXTRA}}
  {\newcommand{\XTRA}[1]{\phantom{}\begingroup\slshape\color{RoyalBlue}\ignorespaces#1\ignorespaces\endgroup}}
  {\newcommand{\XTRA}[1]{}}

\DeclareMathAlphabet{\mathpzc}{OT1}{pzc}{m}{it}

\makeatletter
\newcommand*{\textoverline}[1]{$\overline{\hbox{#1}}\m@th$}
\makeatother

\makeatletter
\@ifclassloaded{IEEEtran}{%
}{%
  
}%

\makeatletter 
\newcommand{\linebreakand}{%
  \end{@IEEEauthorhalign}
  \hfill\mbox{}\par
  \mbox{}\hfill\begin{@IEEEauthorhalign}
}
\makeatother 

\newlist{myReuseInterval}{enumerate}{1}
\setlist[myReuseInterval]{label=(R$_{\arabic*}$),nosep}

\crefname{section}{\S}{\S}
\crefformat{section}{#2\S#1#3}

\crefname{figure}{Fig.}{Figs.}
\Crefname{figure}{Figure}{Figures}

\crefname{equation}{Eq.}{Eqs.}

\definecolor{shadecolor}{gray}{0.9} 

\makeatletter
\@ifpackageloaded{algorithm2e}{\@ifclassloaded{IEEEtran}{%
  \SetAlFnt{\small}
  \SetAlCapFnt{\small}
  \SetAlCapNameFnt{\small}
}}%
\makeatother

\newcommand{\lstsetCommon}{%
  \lstset{%
    columns=fullflexible,%
    keepspaces=true,%
    escapeinside={<[}{]>},%
    moredelim=**[is][\color{Cerulean}]{<*}{*>},
    moredelim=**[is][\color{Red}]{<^}{^>},
    basicstyle=\ttfamily\footnotesize,%
    showstringspaces=false,%
    aboveskip=0em,%
    belowskip=0em,%
  }%
}

\newcommand{\lstsetMe}{%
  \lstsetCommon{}%
  \lstset{%
  }
}

\lstsetMe{}

\usepackage{soul}

\newif\ifhl{}
\hltrue{}
\hlfalse{}

\ifhl{}
 
\else
 
\fi

\newif\ifdraft{}
\drafttrue{}

\ifdraft{}
  \newcommand{\davidnote}[1]{ {\textcolor{purple} { ***[David]: #1 }}}
  \newcommand{\raynote}[1]{ {\textcolor{orange} { ***[Ray]: #1 }}}
  \newcommand{\ozgurnote}[1]{ {\textcolor{blue} { ***[Ozgur]: #1 }}}
  \newcommand{\NOTE}[1]{\phantom{}\begingroup\relax\ifmmode\boldmath\else\bfseries\fi\color{Cerulean}\ignorespaces#1\ignorespaces\endgroup}
  \newcommand{\TODO}[1]{\phantom{}\begingroup\relax\ifmmode\else\sffamily\fi\color{BurntOrange}\ignorespaces#1\ignorespaces\endgroup}
  \newcommand{\FIXME}[1]{\phantom{}\begingroup\relax\ifmmode\boldmath\else\bfseries\sffamily\fi\color{Red}\ignorespaces#1\ignorespaces\endgroup}
  \newcommand{\FIXED}[1]{\phantom{}\begingroup\relax\ifmmode\else\sffamily\fi\color{Green}\ignorespaces#1\ignorespaces\endgroup}
  \newcommand{\REPLACE}[1]{\phantom{}\begingroup\relax\ifmmode\else\sffamily\fi\color{Purple}\ignorespaces#1\ignorespaces\endgroup}
  \newcommand{\DELETE}[1]{\phantom{}\begingroup\relax\ifmmode\else\sffamily\fi\color{Red}\ifmmode\text{\sout{\ensuremath{#1}}}\else\sout{\ignorespaces#1\ignorespaces}\fi\endgroup}
\else
  \newcommand{\davidnote}[1]{}
  \newcommand{\raynote}[1]{}
  \newcommand{\ozgurnote}[1]{}
  \newcommand{\NOTE}[1]{}
  \newcommand{\TODO}[1]{}
  \newcommand{\FIXME}[1]{#1}
  \newcommand{\FIXED}[1]{#1}
  \newcommand{\DELETE}[1]{}
  \newcommand{\REPLACE}[1]{#1}
\fi

\def\BibTeX{{\rm B\kern-.05em{\sc i\kern-.025em b}\kern-.08em
    T\kern-.1667em\lower.7ex\hbox{E}\kern-.125emX}}

\copyrightyear{2025} 
\acmYear{2025} 
\setcopyright{cc}
\setcctype{by-nc}
\acmConference[Accepted at Supercomputing25 PMBS  Workshop]{Workshops of the International Conference for High Performance Computing, Networking, Storage and Analysis}{November 16--21, 2025}{St Louis, MO, USA}
\acmBooktitle{Workshops of the International Conference for High Performance Computing, Networking, Storage and Analysis (SC Workshops '25), November 16--21, 2025, St Louis, MO, USA}
\acmDOI{10.1145/3731599.3769277}
\acmISBN{979-8-4007-1871-7/2025/11}

\begin{document}

\title{CGSim: A Simulation Framework for Large Scale Distributed Computing Environment}

\author{Sairam Sri Vatsavai}
\affiliation{%
  \institution{Brookhaven National Laboratory}
  \city{Upton}\state{NY}\country{USA}
}

\author{Raees Khan}
\affiliation{%
  \institution{University of Pittsburgh}
  \city{Pittsburgh}\state{PA}\country{USA}
}

\author{Kuan-Chieh Hsu}
\affiliation{%
  \institution{Brookhaven National Laboratory}
  \city{Upton}\state{NY}\country{USA}
}

\author{Ozgur O. Kilic}
\affiliation{%
  \institution{Brookhaven National Laboratory}
  \city{Upton}\state{NY}\country{USA}
}

\author{Paul Nilsson}
\affiliation{%
  \institution{Brookhaven National Laboratory}
  \city{Upton}\state{NY}\country{USA}
}

\author{Tatiana Korchuganova}
\affiliation{%
  \institution{University of Pittsburgh}
  \city{Pittsburgh}\state{PA}\country{USA}
}

\author{David K. Park}
\affiliation{%
  \institution{Brookhaven National Laboratory}
  \city{Upton}\state{NY}\country{USA}
}

\author{Sankha Dutta}
\affiliation{%
  \institution{Brookhaven National Laboratory}
  \city{Upton}\state{NY}\country{USA}
}

\author{Yihui Ren}
\affiliation{%
  \institution{Brookhaven National Laboratory}
  \city{Upton}\state{NY}\country{USA}
}

\author{Joseph Boudreau}
\affiliation{%
  \institution{University of Pittsburgh}
  \city{Pittsburgh}\state{PA}\country{USA}
}

\author{Tasnuva Chowdhury}
\affiliation{%
  \institution{Brookhaven National Laboratory}
  \city{Upton}\state{NY}\country{USA}
}

\author{Shengyu Feng}
\affiliation{%
  \institution{Carnegie Mellon University}
  \city{Pittsburgh}\state{PA}\country{USA}
}

\author{Jaehyung Kim}
\affiliation{%
  \institution{Carnegie Mellon University}
  \city{Pittsburgh}\state{PA}\country{USA}
}

\author{Scott Klasky}
\affiliation{%
  \institution{Oak Ridge National Laboratory}
  \city{Oak Ridge}\state{TN}\country{USA}
}

\author{Tadashi Maeno}
\affiliation{%
  \institution{Brookhaven National Laboratory}
  \city{Upton}\state{NY}\country{USA}
}

\author{Verena Ingrid Martinez Outschoorn}
\affiliation{%
  \institution{University of Massachusetts}
  \city{Amherst}\state{MA}\country{USA}
}

\author{Norbert Podhorszki}
\affiliation{%
  \institution{Oak Ridge National Laboratory}
  \city{Oak Ridge}\state{TN}\country{USA}
}

\author{Frédéric Suter}
\affiliation{%
  \institution{Oak Ridge National Laboratory}
  \city{Oak Ridge}\state{TN}\country{USA}
}

\author{Wei Yang}
\affiliation{%
  \institution{SLAC National Accelerator Laboratory}
  \city{Menlo Park}\state{CA}\country{USA}
}

\author{Yiming Yang}
\affiliation{%
  \institution{Carnegie Mellon University}
  \city{Pittsburgh}\state{PA}\country{USA}
}

\author{Shinjae Yoo}
\affiliation{%
  \institution{Brookhaven National Laboratory}
  \city{Upton}\state{NY}\country{USA}
}

\author{Alexei Klimentov}
\affiliation{%
  \institution{Brookhaven National Laboratory}
  \city{Upton}\state{NY}\country{USA}
}

\author{Adolfy Hoisie}
\affiliation{%
  \institution{Brookhaven National Laboratory}
  \city{Upton}\state{NY}\country{USA}
}

\begin{abstract}
Large-scale distributed computing infrastructures such as the Worldwide LHC Computing Grid (WLCG) require comprehensive simulation tools for evaluating performance, testing new algorithms, and optimizing resource allocation strategies. However, existing simulators suffer from limited scalability, hardwired algorithms, lack of real-time monitoring, and inability to generate datasets suitable for modern machine learning approaches. We present CGSim, a simulation framework for large-scale distributed computing environments that addresses these limitations. Built upon the validated SimGrid simulation framework, CGSim provides high-level abstractions for modeling heterogeneous grid environments while maintaining accuracy and scalability. Key features include a modular plugin mechanism for testing custom workflow scheduling and data movement policies, interactive real-time visualization dashboards, and automatic generation of event-level datasets suitable for AI-assisted performance modeling. We demonstrate CGSim's capabilities through a comprehensive evaluation using production ATLAS PanDA workloads, showing significant calibration accuracy improvements across WLCG computing sites. Scalability experiments show near-linear scaling for multi-site simulations, with distributed workloads achieving \xspeed{6} better performance compared to single-site execution. The framework enables researchers to simulate WLCG-scale infrastructures with hundreds of sites and thousands of concurrent jobs within practical time budget constraints on commodity hardware.
\end{abstract}

\maketitle




\section{Introduction}
\label{sec:intro}
\input{introduction}





\section{Related Work}
\label{sec:related}
\input{related}
\section{Simulation Framework}
\label{sec:framework}
\input{implementation}

\section{Evaluation}
\label{sec:evaluation}
\input{evaluation}

\vspace{-0.3cm}
\section{Conclusion}
\label{sec:conclusion}
\input{conclusion}
\vspace{-0.3cm}
\begin{acks}
This material is based on work supported by the U.S. Department of Energy, Office of Science, Office of Advanced Scientific Computing Research under Award Number DE-SC-0012704. This work was done in collaboration with the distributed computing research and development program within the ATLAS Collaboration. We thank our ATLAS colleagues for their support, particularly the ATLAS Distributed Computing team's contributions.
\end{acks}

\vspace{-5pt}
\clearpage
\bibliographystyle{ACM-Reference-Format}
\bibliography{ref.bib}

\end{document}

%% file: introduction.tex
Distributed computing has become the standard for modern data processing, powering applications from global cloud services to large-scale scientific analytics. A notable example is the WLCG, which supports high-energy physics experiments such as ATLAS at the European Organization for Nuclear Research (CERN) \cite{atlas}. The WLCG processes petabytes of data and runs hundreds of thousands of jobs daily to analyze particle collisions from the Large Hadron Collider (LHC) \cite{lhc}. As data volumes and complexity continue to grow, the efficient and reliable operation of such infrastructures is critical.
The performance of distributed systems is typically measured using metrics such as queue time, CPU efficiency, job failure rate, and throughput, all derived from operational logs and monitoring data \cite{kilic2025towards,park2024ai}. These metrics reflect the impact of job scheduling algorithms and data management policies \cite{feng2025alternative}. However, testing new policies directly on production systems is impractical due to scale, risk, and the continuous nature of scientific operations. Although limited trials may occur at individual sites, the prediction of system-wide effects remains infeasible. Simulation, at an appropriate level of abstraction, provides a practical alternative for evaluating and validating new strategies.

Several simulation frameworks have been proposed for distributed computing ~\cite{gridsim, cloudsim, wrench, simgrid, dcsim}. However, only a few specifically target scientific workflows~\cite{wrench, dcsim}, and existing simulators suffer from several shortcomings. First, prior efforts often model only specific components of a system while abstracting others, creating a fidelity gap that prevents the reliable modeling of large-scale infrastructures. Second, scalability remains a challenge, not only in terms of number of workflow tasks but also in simulating platforms with many parallel computing sites. Third, a lack of modularity often prevents researchers from easily plugging in and testing new policies. The core logic is frequently hardwired into the simulator, making it difficult to incorporate and evaluate novel scheduling or data-movement policies. Fourth, the absence of real-time monitoring forces researchers to rely on tedious post-processing of logs to analyze system dynamics. Finally, with the emergence of ML-assisted simulation ~\cite{Elbattah2018}, models need detailed training data sets to act as fast surrogates for performance prediction. Existing simulators rarely provide such statistics, limiting the adoption of modern data-driven methodologies.


To address these shortcomings, we propose the Computing Grid Simulator (CGSim). Built on the open-source SimGrid framework, CGSim leverages well-validated simulation models while providing high-level abstractions for large-scale distributed systems. Unlike prior works that are often limited to a few computing sites due to scalability challenges, CGSim is capable of performing large-scale simulations efficiently. Furthermore, CGSim provides researchers with a flexible and modular plugin mechanism to test their workflow scheduling or data movement policies without modifying the simulator's core code. For detailed analysis, CGSim includes an interactive, web-based dashboard for monitoring simulations in real-time, allowing for the evaluation of system behavior down to the CPU level under new policies. In addition, CGSim automatically generates an event-level statistics dataset from each run that can be directly used to train machine learning models. As a case study, we employed CGSim to simulate the large-scale ATLAS computing grid. The simulator was calibrated using job log records from the PanDA workflow management system \cite{panda} and subsequently validated to ensure fidelity. 
\vspace{-0.3cm}


%% file: related.tex
Simulation has long been a cornerstone of distributed computing research~\cite{zheng2004, Bell2003}. Early frameworks such as GridSim~\cite{gridsim} and CloudSim~\cite{cloudsim} provided accessible environments for modeling grid and cloud systems but often relied on coarse-grained models that limited their accuracy, particularly for data-intensive workloads. The SimGrid~\cite{simgrid} framework addressed these accuracy gaps by introducing a validated discrete-event core and scalable resource-sharing models, establishing a new standard for fidelity. However, this came at the cost of a significant engineering effort for users. To bridge this usability gap, WRENCH~\cite{wrench} was developed as a higher-level framework on top of SimGrid. It provides reusable services (e.g., batch compute, storage, registries) and APIs to enable the implementation of full Workflow Management System (WMS) simulators with modest coding effort. WRENCH also introduced features like lightweight monitoring and a web dashboard for quick inspection of results. Building on this foundation, recent tools have targeted more specific domains. DCSim~\cite{dcsim}, for example, focuses on High-Energy Physics (HEP) infrastructures with features relevant to the CMS experiment \cite{CMS}, such as streaming jobs that pipeline I/O with computation and XRootD-like data caching \cite{dcsim}. DCSim's validation was primarily conducted against controlled testbed traces. Like the aforementioned tools, our proposed simulator, CGSim, is SimGrid-based, but it's architected to address several key gaps we identified in prior work. Specifically, CGSim provides multi-site scalability, a plugin mechanism for policy modularity, advanced real-time monitoring, and automatic dataset generation for ML training. 

\vspace{-0.3cm}

%% file: implementation.tex
\begin{figure*}
    \centering
    \includegraphics[width=\linewidth]{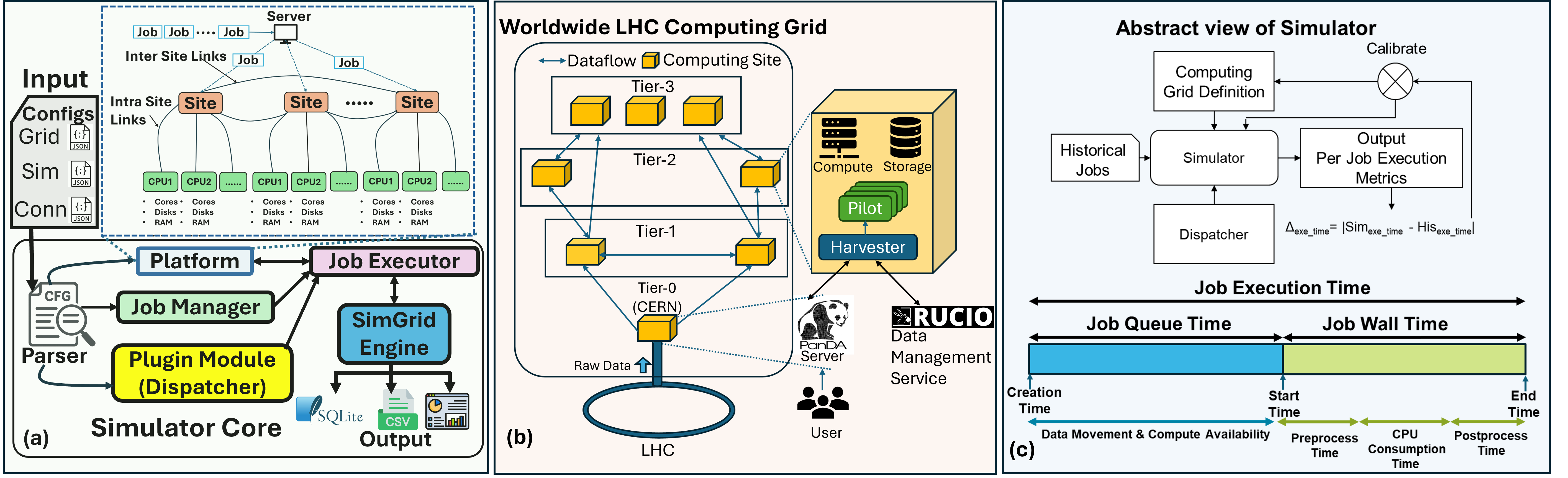}
    \caption{(a) Architecture Overview of CGSim (b) ATLAS computing Grid (c) Calibration methodology.}
    \vspace{-0.5cm}
    \label{fig:simulator}
\end{figure*}
\vspace{-0.1cm}
\subsection{Architecture Overview}
Figure~\ref{fig:simulator}(a) illustrates the overall architecture of CGSim, which employs a layered design to enable accurate modeling of heterogeneous grid environments. The simulator's architecture comprises three distinct layers: input processing, simulation core, and output generation.
 The \textbf{input layer} configures the simulation environment through three JSON files specifying the computational infrastructure, network topology, and execution parameters. This design also enables flexible setup and reproducible experiments. The \textbf{simulation core} interprets inputs, manages hierarchical resources, orchestrates job lifecycles, and executes tasks using the SimGrid discrete-event engine, supporting various scheduling and allocation strategies via a modular plugin system. The \textbf{output layer} collects and stores results in SQLite databases, supports CSV exports for statistical analysis, and provides a real-time dashboard for monitoring and performance evaluation.
\vspace{-0.3cm}
\subsection{Simulation Engine and Core Components}
The simulator is implemented using the SimGrid framework~\cite{simgrid}. The network topology specified in the input configuration initializes the simulated computing grid. As shown in Figure~\ref{fig:simulator}(a), each computing site is modeled as a SimGrid netzone, which acts as a network container to handle routing between its internal hosts and to other netzones. These netzones contain hosts (CPUs) with properties such as speed, RAM, and storage. Sites are interconnected through links that reflect the latency and bandwidth defined in the configuration. An additional host serves as the main server, linked to all sites in the platform. Each site includes a \emph{receiver} actor that retrieves and executes workloads from a local queue, while the main server hosts a \emph{sender} actor, which assigns workloads to the sites by placing them into the respective site queues. This design allows coordinated workload distribution and execution across the simulated network.




The main server acts as the central controller for the simulation, organizing the workflow. On a SimGrid engine run, the main server starts receiving workload information from the job manager, then consults an allocation algorithm (user defined through a plugin) to assign the workload to a resource. Once assigned the main server sends the workload to the assigned sites. If no suitable resource is found, the main server puts the job into a pending list and moves forward with processing the remaining of the incoming workload. Whenever a resource on the grid becomes available, the main server checks the pending list and assigns job accordingly. The simulation finishes once all workloads are assigned and executed. Output metrics are dumped periodically to be post processed for analysis.
\begin{figure}
    \centering
    \includegraphics[width=\linewidth]{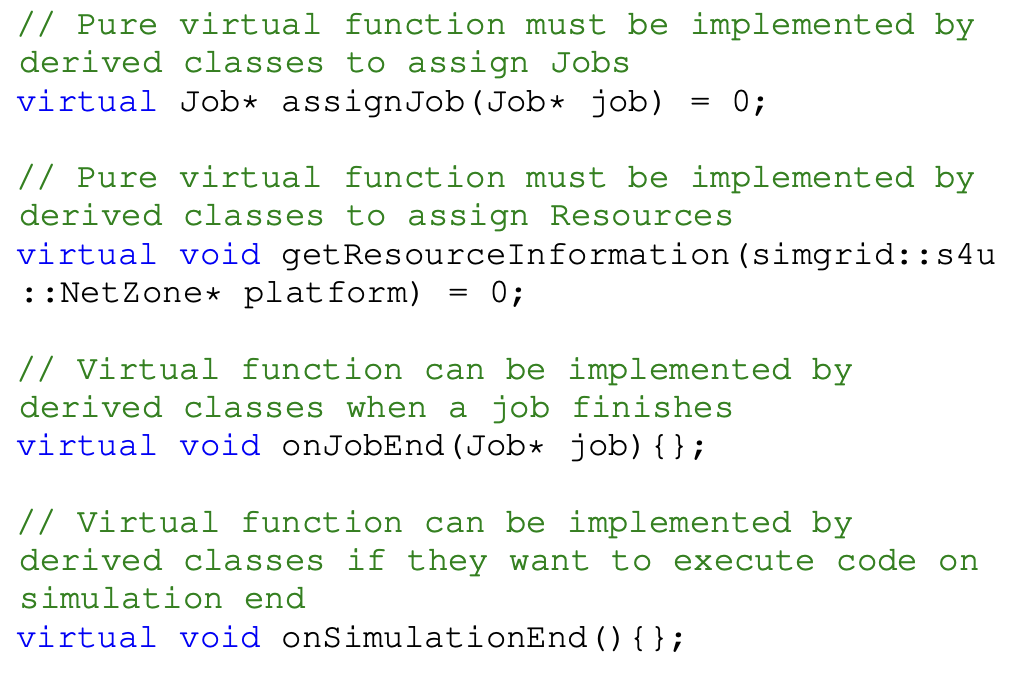}
    \caption{Abstract class to allow users to define their own allocation policies.}
    \vspace{-0.5cm}
    \label{fig:plugin-code}
\end{figure}
\vspace{-0.3cm}
\subsection{Plugins}
One of the main features of CGSim is to allow users to easily test custom workload allocation algorithms through a plugin system. Plugins containing user defined code are implemented as shared libraries that can be built independently and loaded into the simulation via the input configuration. This design allows users to incorporate custom algorithms without modifying the simulator’s core code.

To facilitate plugin development, CGSim provides an abstract class which serves as a blueprint for writing plugins. The abstract class is automatically installed on installation of the CGSim package. User-defined plugin classes can inherit from this base class and override its functions to implement their custom algorithms. The functions users can override are shown in Figure~\ref{fig:plugin-code}. They serve as hooks to allow the algorithm to communicate its strategy with the simulation. 

\texttt{assignJob} is the main method which must be implemented by the user with a custom allocation strategy. CGSim uses a standardized job (workload) structure, which is installed as a header. The goal of the user is to assign the \emph{allocation site} field for all incoming jobs using information about the workload contained in the job structure and resource information that must be configured in the plugin via the \texttt{getResourceInformation} method which provides the user access to the grid topology (platform) defined in SimGrid.

CGSim comes equipped with a simple plugin example that can be built and used out of the box. It also serves as a platform for the development of plugins containing more complicated allocation strategies. Detailed instructions and a tutorial on how to write plugins can be found on the project website.

%% file: evaluation.tex
\vspace{-0.1cm}
\subsection{Case Study: ATLAS Computing Grid}
As a case study, we simulate the the subset of the WLCG that supports the ATLAS experiment shown in Figure ~\ref{fig:simulator}(b). ATLAS  is a high-energy physics experiment at the LHC located at CERN in Geneva, Switzerland. The ATLAS detector investigates particle collisions at high energies, generating petabytes of data annually in the search for new physics discoveries. Thousands of scientists across the world analyze this massive dataset remotely using the ATLAS globally distributed computing infrastructure which spans approximately 200 computing centers across more than 40 countries. This distributed analysis ecosystem relies on two critical systems: PanDA for workload management and Rucio \cite{rucio} for data management, which together coordinate the complex computational demands of modern particle physics research. Figure~\ref{fig:simulator}(b) also illustrates the architectural flow of these two systems across the WLCG.

For our simulation study, we model the WLCG topology and resource characteristics using CGSim, focusing on the computational aspects of job execution across distributed sites. We configure the simulator with realistic site capacities, network topologies, and workload patterns derived from operational ATLAS computing metrics. 

\begin{figure}
    \centering
    \includegraphics[width=\linewidth]{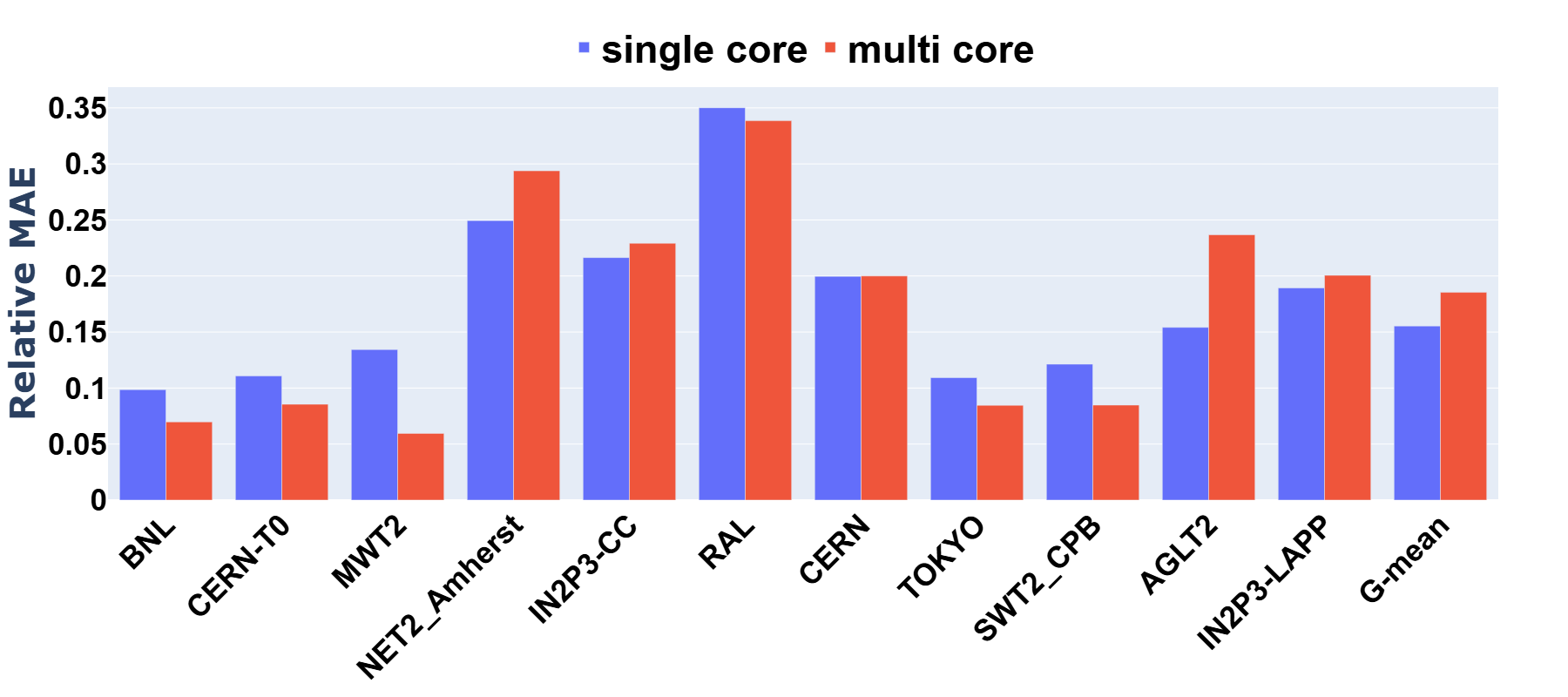}
    \caption{Job walltime calibration of CGSim for single-core and multi-core jobs across the 50 sites of WLCG. For brevity, we are plotting only 10 sites. Geometric mean is computed for all the sites.}
    \vspace{-0.5cm}
    \label{fig:calibration}
\end{figure}
\vspace{-0.3cm}
\subsection{Calibration and Validation} \label{calibrationsec}

Figure~\ref{fig:simulator}(c) illustrates our calibration methodology for ensuring simulation accuracy against the real-world ATLAS computing infrastructure. We establish the ATLAS grid topology within CGSim using site configuration parameters derived from HEPScore23 benchmarking data of WLCG computing centers \cite{szczepanek2024hep}.\textbf{Data Collection and Preprocessing:} We collect historical job execution records from the PanDA workload management system, containing production workloads over a 6-month time period (January 2024 - June 2024). Each preprocessed job record contains essential characteristics including computational requirements, timestamp information, input/output file counts, and target computing site assignments. Critically, each record provides ground truth measurements for total job execution time, decomposed into job walltime (actual processing duration) and job queue time (scheduling and resource allocation delays). \textbf{Calibration Framework:} Our calibration process follows PanDA's dispatching policies to replicate realistic job to site assignments. We perform site specific calibration by feeding historical jobs into the simulator and measuring the discrepancy between ground truth execution time ($His_{\text{exe\_time}}$) and simulated execution time ($Sim_{\text{exe\_time}}$). The calibration objective is to minimize $\Delta exe_\text{time}$ = $Sim_{\text{exe\_time}} - His_\text{exe\_time}$ across all sites and job types.

We address two primary sources of simulation error: (1) implementation gaps within the simulation core that require algorithmic corrections, and (2) configuration parameter misalignment that can be resolved through systematic tuning. While implementation gap identification represents an iterative refinement process, we focus on systematic parameter calibration for quantifiable accuracy improvements.
\textbf{Parameter Sensitivity Analysis:} Through comprehensive sensitivity analysis, we evaluate the impact of various grid configuration parameters on job execution accuracy, including CPU core counts, processing speeds, memory capacities, and intra-site network bandwidths. Our analysis identifies CPU core processing speed as the dominant factor influencing job walltime accuracy, establishing it as the primary calibration parameter for each computing site.
\textbf{Optimization Methods:} We evaluate four calibration approaches: brute force search, random sampling, Bayesian optimization (BO), and Covariance Matrix Adaptation Evolution Strategy (CMA-ES) \cite{hansen2016cma}. Brute-force search is theoretically optimal but computationally infeasible across 150 sites. 
Among the methods, random search demonstrates superior performance, achieving the lowest average error across 50 computing sites, likely due to the parameter optimization landscape. Figure ~\ref{fig:calibration} shows the relative mean absolute error obtained by the random search calibration across the 50 sites. Our calibration improved the geometric mean of relative mean absolute error of single core and multi core jobs from 76\% to 17\% across 50 sites.   
Following walltime calibration, we extend the methodology to queue time modeling, incorporating scheduling overhead and resource contention effects to achieve comprehensive job lifecycle accuracy.
\begin{figure}
    \centering
    \includegraphics[width=\linewidth]{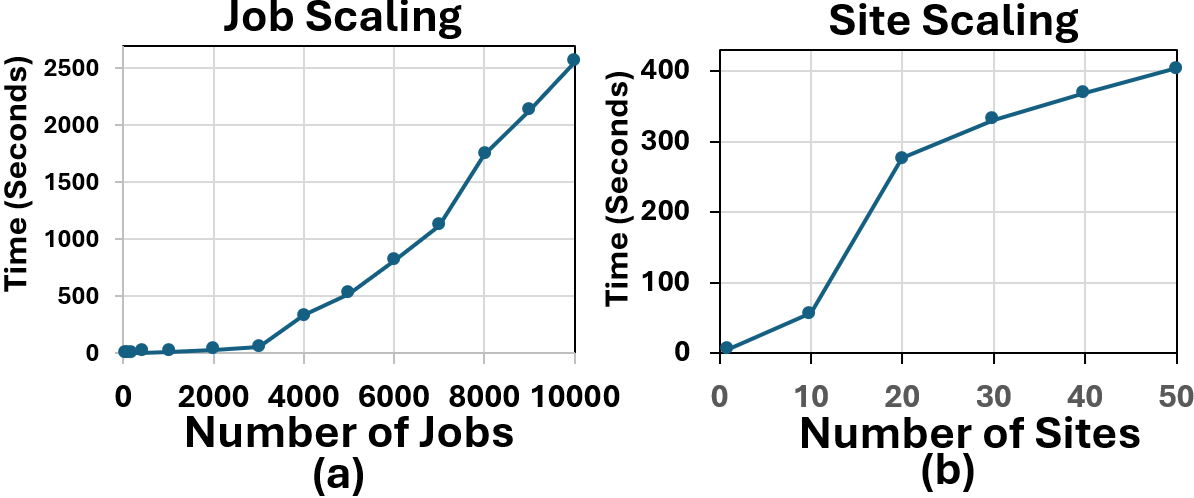}
    \caption{Scalability analysis of CGSim: (a) job scaling performance with increasing workload density per single site, and (b) multi-site scaling performance with fixed job density across 1-50 computing sites.}
    \vspace{-0.5cm}
    \label{fig:scalibility}
\end{figure}
\vspace{-0.3cm}
\subsection{Experimental Results}
\subsubsection{Scalability}
To evaluate CGSim's scalability for large-scale distributed computing simulations, we conducted experiments along two critical dimensions: job scaling (i.e., increasing number of jobs per site) and multi-site scaling (i.e., increasing number of sites). We perform scalability experiments using historical PanDA job records collected from production WLCG operation (see Section \ref{calibrationsec}). For multi-site scaling, we distribute a fixed workload of PanDA job records across an increasing number of sites, ranging from 1 to 50 sites while maintaining 200 jobs per site. Each site was configured with 100–2,000 CPU cores, consistent with actual WLCG site specifications. Experiments are executed on a system equipped with Intel Core i9-13900H and 64GB RAM, with multiple runs per configuration to ensure statistical correctness.

Figure \ref{fig:scalibility}(a) demonstrates CGSim's job-scaling performance as workload density increases. The simulator exhibits sub-quadratic scaling characteristics, with execution time growing from under 100 seconds for 1,000 jobs to approximately 2,500 seconds for 10,000 jobs. This scaling behavior indicates that CGSim can efficiently handle job densities equivalent to those observed during peak ATLAS data processing periods without experiencing prohibitive runtime growth.

Figure \ref{fig:scalibility}(b) illustrates multi-site scaling results, showing near-linear scaling behavior as the number of simulated sites increases from 1 to 50. Simulation runtime grows from under 50 seconds for single-site configurations to approximately 400 seconds for 50-site scenarios. This linear scaling validates CGSim's architectural design for distributed simulation and demonstrates its capability to model WLCG-scale infrastructures comprising hundreds of computing centers.


\begin{table}[htbp]
\caption{Representative sample of event-level monitoring data captured by CGSim.}
\label{table:event_monitoring}
\centering
\resizebox{\columnwidth}{!}{%
\begin{tabular}{|c|c|c|c|c|c|c|c|}
\hline
\multicolumn{1}{|c|}{\textbf{\begin{tabular}[c]{@{}c@{}}Event\\ ID\end{tabular}}} & \multicolumn{1}{c|}{\textbf{Job ID}} & \multicolumn{1}{c|}{\textbf{State}} & \multicolumn{1}{c|}{\textbf{Site}} & \multicolumn{1}{c|}{\textbf{\begin{tabular}[c]{@{}c@{}}Avail.\\ Cores\end{tabular}}} & \multicolumn{1}{c|}{\textbf{\begin{tabular}[c]{@{}c@{}}Pending \\ Jobs\end{tabular}}} & \multicolumn{1}{c|}{\textbf{\begin{tabular}[c]{@{}c@{}}Assigned\\ Jobs\end{tabular}}} & \multicolumn{1}{c|}{\textbf{\begin{tabular}[c]{@{}c@{}}Finished\\ Jobs\end{tabular}}} \\ \hline
8570 & 6466065355 & finished & DESY-ZN & 66120 & 0 & 134 & 62 \\ \hline
8571 & 6465869354 & finished & LRZ-LMU & 39998 & 0 & 95 & 100 \\ \hline
8573 & 6471661661 & finished & BNL & 74330 & 0 & 157 & 37 \\ \hline
8577 & 6466259044 & finished & CERN & 39504 & 0 & 118 & 79 \\ \hline
\end{tabular}%
}
\end{table}
\vspace{-0.3cm}

\subsubsection{Event Level Simulation Snapshot}
CGSim captures comprehensive simulation state at each timestep, recording both job-level transitions and site-level resource dynamics. As illustrated in Table~\ref{table:event_monitoring}, the monitoring system tracks individual job states (pending, assigned, running, finished, failed) with precise timestamps , alongside concurrent site metrics including available cores, job queue depths, and cumulative completion statistics.
This dual-level tracking enables detailed analysis of system behavior over time, revealing how individual job scheduling decisions impact overall site utilization and queue dynamics. The structured output format supports both real time monitoring during simulation execution and post-processing for performance analysis and machine learning dataset generation.

\subsubsection{Visualization and Monitoring}
\begin{figure}
    \centering
    \includegraphics[width=\linewidth]{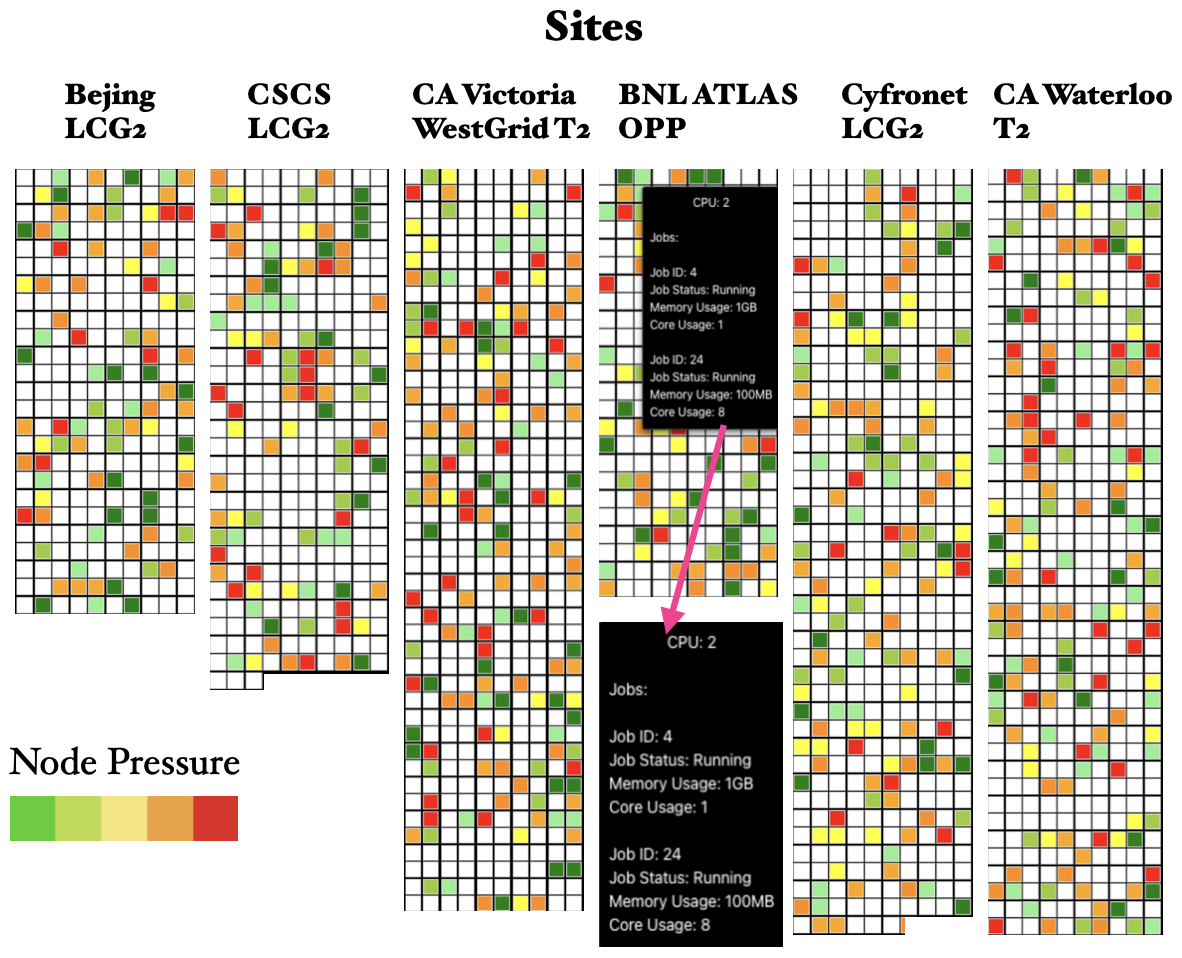}
    \caption{Monitoring provides a visual analysis of the workload distribution, with hover-over details showing the jobs running on each node.}
    \vspace{-0.5cm}
    \label{fig:monitoring}
\end{figure}

CGSim provides an interactive real-time dashboard that visualizes the operational state of all simulated computing sites simultaneously as shown in Figure~\ref{fig:monitoring}.
The dashboard enables detailed inspection of individual jobs by displaying their ID, execution status, memory usage, and core allocation, while its multi-site visualization reveals system-wide behavior and real-time load distribution. The node pressure shows the number of CPUs being utilized at each site, providing immediate visibility into computational load across the distributed infrastructure. This allows one to detect resource bottlenecks, monitor infrastructure performance, and gain deeper insights into workload dynamics. This functionality enables users to detect resource bottlenecks, monitor infrastructure performance, and gain deeper insight into workload dynamics. The interactive interface complements structured data exports by supporting dynamic monitoring during simulations, allowing immediate identification of scheduling inefficiencies, contention hotspots, and emergent load-balancing behaviors. These insights are often difficult to extract from post-processed aggregate statistics alone.

%% file: conclusion.tex
We presented CGSim, a simulation framework that addresses critical limitations of existing distributed computing simulators by combining SimGrid's validated models with high-level abstractions, a modular plugin architecture, and real-time monitoring capabilities. This design enables scalable, multi-site modeling of large-scale infrastructures such as the WLCG. Through comprehensive evaluation using ATLAS PanDA records, we demonstrated substantial accuracy improvements and near-linear scaling characteristics that validate CGSim's ability to model hundreds of computing centers with thousands of concurrent jobs. This combination of production-scale fidelity, algorithmic modularity, and comprehensive observability enables researchers to safely evaluate infrastructure designs, test novel scheduling algorithms, and optimize resource allocation strategies without the risks and costs of production system experimentation. Our future work will focus on extending scalability to full WLCG scale, comprehensive comparison with existing simulators, and integrating advanced machine learning techniques for automated calibration and surrogate modeling.